\documentclass[iop]{emulateapj}
\usepackage{natbib}
\usepackage{comment}
\usepackage{amsmath}
\usepackage{hyperref,breakurl}

\shorttitle{Microwave Continuum Emission and Dense Gas Tracers in NGC\,3627}
\shortauthors{MURPHY ET AL.}
\slugcomment{Accepted to ApJ on \today}

\begin{document}
\title{Microwave Continuum Emission and Dense Gas Tracers in NGC\,3627: Combining Jansky VLA and ALMA Observations}

\author{Eric J. Murphy\altaffilmark{1}, Dillon Dong\altaffilmark{2}, Adam K. Leroy\altaffilmark{3}, Emmanuel Momjian\altaffilmark{4}, 
James J. Condon\altaffilmark{5}, George Helou\altaffilmark{6}, David S. Meier\altaffilmark{7,4}, J\"{u}rgen Ott\altaffilmark{4}, Eva Schinnerer\altaffilmark{8}, Jean L. Turner\altaffilmark{9}}

\altaffiltext{1}{Infrared Processing and Analysis Center, California Institute of Technology, MC 314-6, Pasadena, CA 91125, USA; \email{emurphy@ipac.caltech.edu}}
\altaffiltext{2}{Department of Physics and Astronomy, Pomona College, Claremont, CA 91711, USA}
\altaffiltext{3}{The Ohio State University, 140 W18th St, Columbus, OH 43210, USA}
\altaffiltext{4}{National Radio Astronomy Observatory, P.O. Box O, 1003 Lopezville Road, Socorro, NM 87801, USA}
\altaffiltext{5}{National Radio Astronomy Observatory, 520 Edgemont Road, Charlottesville, VA 22903, USA}
\altaffiltext{6}{California Institute of Technology, MC 100-22, Pasadena, CA 91125, USA}
\altaffiltext{7}{New Mexico Institute of Mining \& Technology, 801 Leroy Place, Socorro, NM 87801, USA}
\altaffiltext{8}{Max Planck Institut f\"{u}r Astronomie, K\"{o}nigstuhl 17, Heidelberg D-69117, Germany}
\altaffiltext{9}{Department of Physics and Astronomy, UCLA, Los Angeles, CA 90095, USA}

%\affil{Infrared Processing and Analysis Center, California Institute of Technology, MC 314-6, Pasadena, CA 91125, USA; emurphy@ipac.caltech.edu}
%\afil{}

\begin{abstract}
We present Karl G. Jansky Very Large Array (VLA) Ka band (33\,GHz) and
Atacama Large Millimeter Array (ALMA) Band 3 (94.5\,GHz) continuum %JC deleted hyphen in Band 3
images covering the nucleus and two extranuclear star-forming regions %JC replaced 'imaging' with 'images' because we a presenting images; 'imaging' is just the process of creating images.  I made this change throughout the paper as needed. Hyphenated 'star-forming'
within the nearby galaxy NGC\,3627 (M\,66), observed as part of the
Star Formation in Radio Survey (SFRS).  % Inserted 'SFRS' to make it clear this is the name of a well-defined project.
Both images %JC images, not 'data sets' have angular resolution
achieve an angular 
resolution of $\lesssim$2\arcsec, allowing us to map the radio
spectral indices and estimate thermal radio fractions at a linear %JC replaced ambiguous 'physical' with precise 'linear' 
resolution of $\lesssim$90\,pc at the distance of NGC\,3627.  The
thermal fraction at 33\,GHz reaches unity at and around the peaks of
each H{\sc ii} region; we additionally observed %JC changed 'observe' to past tense
the spectral index between 33 and 94.5\,GHz to become both increasingly negative and positive 
%negative and positive spectral steepening between 33 and 94.5\,GHz %JC sign ambiguity: is 'negative spectral steepening' a spectrum that flattens at high frequencies or steepens at high frequencies?  It appears to be the latter in the remainder of this sentence.  I think of (positive) spectral steepening as meaning steeper, not flatter.
away from the peaks of the H{\sc ii} regions, indicating an increase of non-thermal extended %JC replaced 'diffuse'
emission from diffusing cosmic-ray electrons and the possible presence of cold dust, respectively.  
While the ALMA observations were optimized for collecting continuum data, they also detected %JC replaces 'reveal detections' by 'detected' for brevity.  Also note that I switched from present tense to past tense to describe events that have already happened.  I changed tenses throughout the paper, as needed.
line emission from %JC replaced  'for' by 'from' 
the $J=1\rightarrow0$ transitions of HCN and HCO$^{+}$.
%We additionally find that 
The peaks of dense molecular gas traced by these two spectral lines are spatially offset from the peaks of the 33 and 94.5\,GHz continuum emission for the case of the extranuclear star-forming regions, indicating that our data reach an angular resolution at which one can spatially distinguish sites of recent %JC recent replacs 'current'
star formation from the sites of future %JC simpler parallel than 'for the next-generation of'
star formation.  %JC The last sentence in the abstract seems to be the main conclusion of our paper.
Finally, we find trends of decreasing dense gas fraction and velocity dispersion with increasing star formation efficiency among the three regions observed, indicating that the dynamical state of the dense gas, rather than its abundance, plays a more significant role in the star formation process.

%At these microwave frequencies, emission is powered primarily by free-free emission from young massive stars

\end{abstract}
\keywords{galaxies: individual (NGC\,3627) -- galaxies: star formation -- H{\sc ii} regions -- ISM: molecules -- radio continuum: general}

\section{Introduction}

%Stars form out of small pockets of dense molecular gas and the young, massive stars that are unambiguously young are often embedded behind layers of dust.
Stars form out of small pockets of dense molecular gas, with massive stars that are unambiguously young often embedded behind thick layers of dust.  
The dense star-forming gas is selectively traced by transitions with high critical densities, 
%and low excitation temperature requirements, 
like the $J=1\rightarrow0$ transitions of HCN or HCO$^{+}$ (i.e., $n_{\rm crit} \gtrsim 10^{5}$\,cm$^{-3}$). 
Likwise, recently formed stars are best studied by tracers such as free-free continuum at $\lambda \lesssim 1\,\mathrm{cm}$ or hydrogen recombination-line
tracers that are robust against extinction and selectively sensitive to the massive young stars that live only a short time after their birth. %JC rephrased and shortened previous two sentences
To constrain %JC tracers can 'constrain' but not 'understand'
the complete physical process of star formation, ideally both tracers would be observed on scales matched to individual star-forming regions. However, because of the faintness of both types of emission and the need for wide frequency coverage to isolate free-free emission from
contaminants, this sort of study has been challenging before the current generation of radio and millimeter-wave telescopes, especially the Karl G. Jansky Very Large Array (VLA) and the Atacama Large Millimeter/submillimeter Array (ALMA). %JC  switched VLA and ALMA to match 'radio' and 'mm', respectively.
In this paper, we combined %JC past tense
 new observations from ALMA and the VLA to report one of the first extragalactic cloud-scale comparisons of %JC of replaced 'between'
 dense gas, traced by the high effective density transitions HCN ($J=1\rightarrow0$) and HCO$^+$ ($J=1\rightarrow0$), and recent star formation traced by $\lambda \lesssim 1\,\mathrm{cm}$ %JC The term 'microwave' can mean frequencies as low as 300 MHz (lambda = 1 m), so it is better to specify a shorter wavelength limit to ensure a low free-free optical depth.
free-free emission.

\begin{deluxetable*}{lcccccc}
\tablecaption{Source Names and Imaging Summary \label{tbl-1}}
\tabletypesize{\scriptsize}
\tablewidth{0pt}
\tablehead{
\colhead{Source}  & \colhead{R.A.} & \colhead{Decl.} & \colhead{Beam (33\,GHz)} & \colhead{$\sigma_{\rm 33\,GHz}$} & \colhead{Beam (94.5\,GHz)} & \colhead{$\sigma_{\rm 94.5\,GHz}$}\\
\colhead{} & \colhead{(J2000)} & \colhead{(J2000)} &  & \colhead{($\mu$Jy\,bm$^{-1}$)} &  & \colhead{($\mu$Jy\,bm$^{-1}$)}
}
           NGC\,3627   &$   11   ~20   ~15.0   $&$+   12   ~59   ~30$  &            $2\farcs83 \times 1\farcs83$  & 26  &            $1\farcs89 \times 1\farcs66$  & 32\\
  NGC\,3627~Enuc.\,1   &$   11   ~20   ~16.2   $&$+   11   ~20   ~16$  &            $2\farcs45 \times 2\farcs03$  & 20  &            $1\farcs88 \times 1\farcs67$  & 32\\
  NGC\,3627~Enuc.\,2   &$   11   ~20   ~16.3   $&$+   11   ~20   ~16$  &            $2\farcs55 \times 2\farcs08$  & 25  &            $1\farcs89 \times 1\farcs67$  & 32  
\enddata
\end{deluxetable*}

This paper represents the first results of a larger project to combine ALMA and the VLA to study continuum emission from nearby galaxies at frequencies ranging between $3-100$\,GHz.
Emission at %JC deleted 'microwave'
 frequencies spanning $30-100$\,GHz is expected to be dominated by free-free emission from 
H{\sc ii} regions, %JC replaced 'young massive star-forming regions,'
 providing a highly robust measure of massive star formation activity unbiased by dust.  
Unfortunately, this emission component is energetically weak, making detections difficult and time consuming even for bright objects.  
To date, observations in this frequency range have been largely limited to Galactic H{\sc ii} regions \citep[e.g.,][]{pgm67a}, nearby dwarf irregular 
galaxies \citep[e.g.,][]{kg86}, galaxy nuclei \citep[e.g.,][]{th83,th94}, nearby starbursts \citep[e.g.,][]{kwm88,th85}, and super star clusters within nearby blue compact dwarfs \citep[e.g.,][]{thb98,kj99}.  
However, with ALMA now online, 3\,mm bolometer arrays such as MUSTANG \citep{mustang}, along with recent improvements to the backends of existing radio telescopes, such as the Caltech Continuum Backend (CCB) on the Robert C. Byrd Green Bank Telescope (GBT) and the Wideband Interferometric Digital ARchitecture (WIDAR) correlator on the VLA, the availability of increased bandwidth is making it possible to conduct investigations for larger samples of extragalactic objects at frequencies $\gtrsim$30\,GHz \citep[e.g.,][]{ejm11b,ejm12b,nb12}.  

Taking advantage of this new capability, the Star Formation in Radio Survey \citep[SFRS; see][]{ejm12b}, targets 118 star-forming regions (56 nuclear and 62 extranuclear) %JC moved 'regions' before ()
 in 56 nearby galaxies ($d < 30$\,Mpc) that had been %JC replaced 'were' with 'had been'
 observed at infrared (IR) and optical wavelengths as part of the SINGS \citep{rck03} and KINGFISH \citep{kf11} legacy programs.
Of these, 112 (50 nuclei and 62 extranuclear regions) have  $\delta > -35\degr$ and are thus observable with the VLA, while 54 are observable with ALMA (i.e., $\delta < 30\degr$). %JC rephased previous sentence
We observed 9 of these regions at frequencies near 95\,GHz using ALMA's Band 3 receiver during the second early science campaign (``Cycle 1''). 
All 9 of these targets had already been observed by the VLA. %JC 'All of these targets' is ambiguous.  All 112? All 54?  All 9?

Thanks to ALMA's strong multiplexing capabilities and excellent sensitivity, we were also able to make a simultaneous search for bright line emission near 95\,GHz. Specifically, the bandpass
always covers the rest-frame $J=1\rightarrow0$ transition of HCO$^+$ and often includes the $J=1\rightarrow0$ transition of HCN. %JC added 'rest frame'
These are two of the standard extragalactic tracers of dense molecular gas
\citep[][]{gao04,graciacarpio06}. Thus our combined VLA and ALMA observations yield an unbiased view of the ionizing radiation emitted by massive young stars and
a snapshot of the kind of bright, dense structure that might give birth to these regions.

In this paper, we report on the combination of VLA and ALMA data for three regions in the nearby galaxy NGC\,3627 (M66), having a kinematic local standard of rest velocity $v_\mathrm{LSRK} = 726 \pm 3 \mathrm{km\,s}^{-1}$.
%heliocentric velocity $v_\mathrm{h} = 727 \pm 3 \mathrm{km~s}^{-1}$. %JC specify v_h for the later comparison with the 750 km/s ALMA cutoff
Along with NGC\,3628 and NGC\,3623, 
NGC\,3627 makes up the well-known Leo Triplet galaxy group.  It is classified with a morphological type of SABb\footnote[1]{Kinematic LSR velocity and morphological type taken from the NASA  Extragalactic Database (NED; http://nedwww.ipac.caltech.edu).}, located at a distance of $9.38 \pm 0.35$\,Mpc \citep{hkp01}, and hosts a Seyfert\,2 AGN \citep{hfs97, jm10}. %JC replaced 'is known to host an AGN \citep[i.e., Seyfert\,2;][]{hfs97, jm10}.'
An H{\sc i} plume extending $\approx$50\arcmin~indicates %JC replaced 'A long H{\sc i} plume extending $\approx$50\arcmin~ lends substantial evidence'
that the two largest spirals in the group, NGC\,3627 and 3628, have interacted in the past \citep[e.g.,][]{mh79}.  
%NGC\,3627 has an H$_{2}$ \citep{kuno07} to H{\sc i} \citep{haan08} mass ratio of $\approx$5.1, which is rather high compared to similar galaxies \citep[i.e., $M_{\rm H_{2}}/M_{\rm H\textsc{i}}$ = 0.9;][]{vc04}.  
NGC\,3627 has comparable amounts of atomic and molecular gas \citep{bimasong,things08}, which is a rather high molecular gas fraction compared to other local star-forming galaxies \citep[e.g.,][]{saintonge11}.  
It has been suggested that the high H$_{2}$/H{\sc i} mass ratio in NGC\,3627 is the result of the tidal interaction with NGC\,3628, since this galaxy has stripped much of the H{\sc i} originally %JC inserted 'originally'
in NGC\,3627 \citep{xz93}.  
%In the rest of the paper, we describe the data and analysis procedures used in the present study (\S2), present our results with their implications discussed in \S3. In \S4, we summarize our main conclusions.  

The paper is organized as follow:  
In \S2 we describe the data as well as the analysis procedures used in the present study.  
In \S3 we present our results and discuss their implications.  
Our main conclusions are then summarized in \S4.

\begin{figure}[!t]
\epsscale{1.1}
\plotone{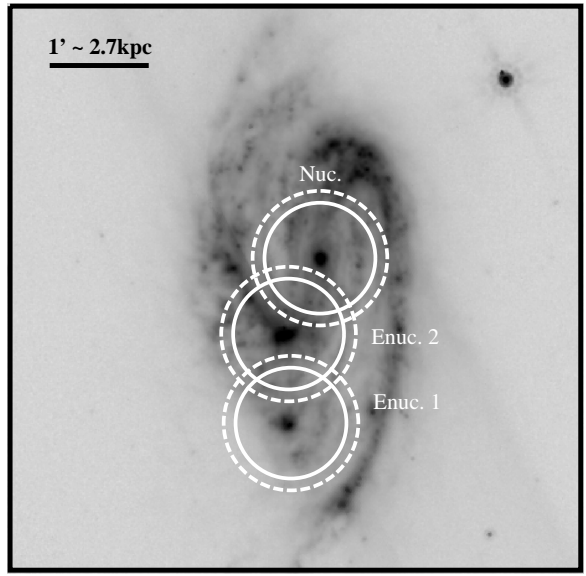}
\caption{The locations of the VLA (dashed line) and ALMA (solid line) pointings for each region in NGC\,3627 overlaid on a %JC inserted space after 'a'
{\it Spitzer} $\lambda = 8\,\mu$m %JC inserted '\lambda = ' because the image is not 8 microns in size
grayscale image.  
The circles outline the FWHMs %JC replace 'The diameters of each circle correspond to the FWHM' 
of the VLA and ALMA primary beams at 33 and 94.5\,GHz, respectively.  
}
\label{fig:pointings}
\end{figure}

\begin{figure*}
\epsscale{1.1}
\plotone{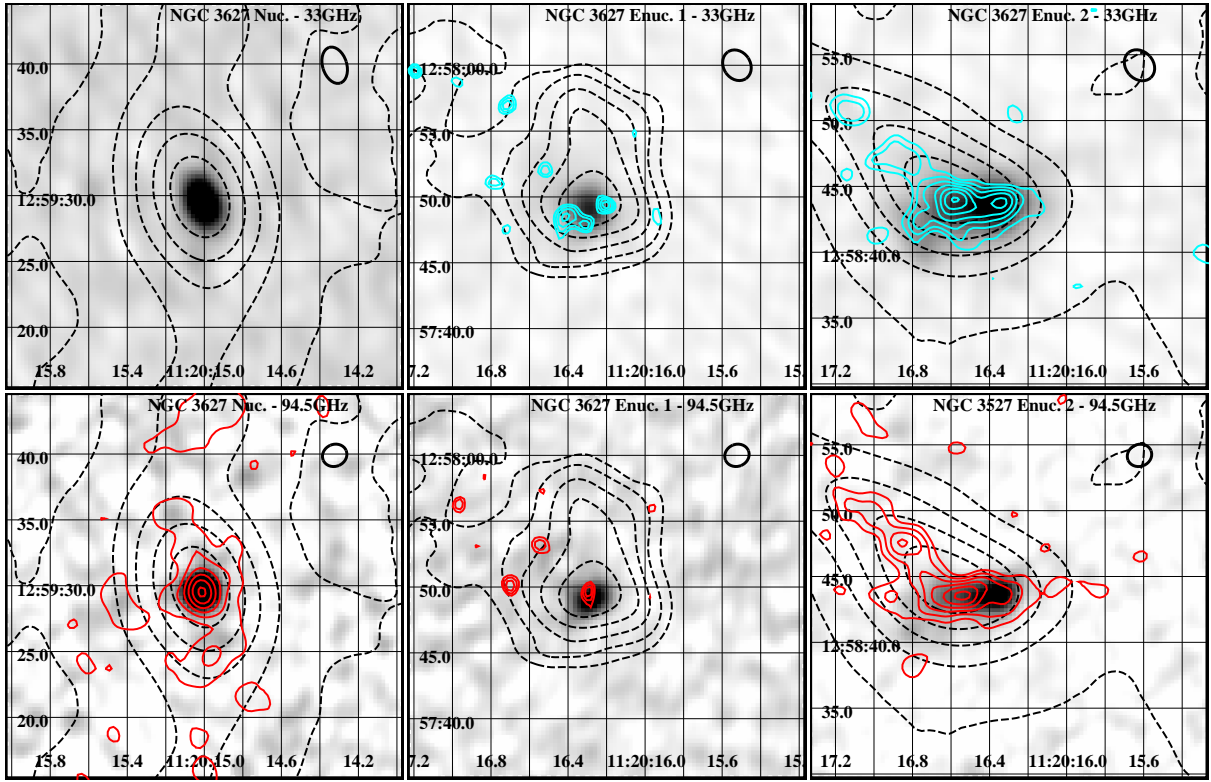}
\caption{Continuum images of the three regions at 33 and 94.5\,GHz %JC replaced 'for each region' 
are shown in the first and second rows, respectively.  
Coordinates are in J2000. 
The restoring %JC deleted the redundant 'clean'
 beam size (FWHM) is illustrated in the upper right corner of each panel.
At the distance of NGC\,3627, 1\arcsec~linearly projects to $\approx$45.5\,pc on the sky.  
CO contours are overlaid on each greyscale image as black dashed lines.  %JC What are the CO contour levels? sigma was not specified for the CO images.
The resolution of the CO images %JC images, not 'data'
 is far lower than in our ALMA or VLA images, %JC likewise for ALMA and VLA images
being $7\farcs3 \times 5\farcs8$.  %pa =  8.019E+01 
HCN contours (cyan) are overlaid on the 33\,GHz grayscale images in the {\it top} three panels, while HCO$^{+}$ contours (red) are overlaid on the 94.5\,GHz grayscale images in the {\it bottom} three panels.  
In all cases, contour lines are drawn using a linear scaling starting from the 2$\sigma$ value.  
As discussed in the text, the full velocity range of HCN emission towards the nucleus is cutoff by the bandpass at $v_\mathrm{LSRK} \approx 800\mathrm{\,km\,s^{-1}}$, and therefore not shown.}
\label{fig:contimgs}
\end{figure*}

\section{Data and Analysis}
We targeted the nucleus and two %JC replaced '2'
 extranuclear star-forming regions in NGC\,3627 with both ALMA and the VLA (Table \ref{tbl-1}), achieving a common resolution of $\approx$2\arcsec,  
% (Table \ref{tbl-1}), 
 which is $\approx$90\,pc at the distance of NGC\,3627. 
In Figure \ref{fig:pointings} we identify the locations of each region targeted in NGC\,3627, showing the full width at half maximum (FWHM) of the VLA and ALMA primary beams at 33 and 94.5\,GHz on a 
{\it Spitzer} $\lambda = 8\,\mu$m %JC inserted 'lambda = '
greyscale image. 
The fields were %JC past tense
 selected to target individual massive star-forming regions, %JC inserted comma
 and the $\lambda = 8\,\mu$m %JC inserted 'lambda = '
image shows that each field is centered on a substantial concentration of bright polycyclic aromatic hydrocarbon (PAH) emission, indicating a combination of intense radiation and abundant small dust grains. 
For the case of NGC\,3627, this yields a sample that includes the nuclear disk, an isolated H{\sc ii} region (Enuc.\,1), and the star-forming complex at the end of a bar (Enuc.\,2), allowing us to compare star formation and molecular gas properties for three very different environmental conditions.  

We worked %JC past tense
 with four data sets: VLA continuum observations at Ka band (29--37\,GHz), ALMA continuum observations centered at $\approx$95\,GHz, ALMA line observations of the $J=1\rightarrow0$ transition of HCO$^+$ and HCN, and archival BIMA Survey of Nearby Galaxies \citep[SONG][]{bimasong} line observations of the $J=1\rightarrow0$ transition of CO.

\subsection{VLA Ka-Band Data}

Details on all of the %JC replaced 'the entire'
SFRS VLA Ka-band survey observations and data reduction can be found in E.J. Murphy et al. (2015, in preparation). Here, we present data only %JC moved 'only'
for NGC\,3627. D-configuration observations were obtained in November 2011 (VLA/11B-032) and March 2013 (VLA/13A-129).
For the first round of observations, the 8-bit samplers were available, yielding 2\,GHz of 
simultaneous %JC replaced 'simultaneous' with 'instantaneous'
bandwidth, which we used to center 1\,GHz wide sub-bands %JC replaced 'spectral basebands'
 at 32.5 and 33.5\,GHz. 
For the latter run, the 3-bit samplers became available, yielding 8\,GHz of 
instantaneous %JC replaced 'simultaneous'
 bandwidth in  2\,GHz wide sub-bands centered at 30, 32, 34, and 36\,GHz. %JC rephrased as above
In both cases, 3C\,286 was used as the flux density and bandpass calibrator, while J\,1118+1234 was used as the complex gain and telescope pointing calibrator. %JC shortened sentence to eliminate 'and to periodically correct the telescope pointing.'
To reduce the VLA data, we used the Common Astronomy Software Applications \citep[CASA;][]{casa} and followed standard procedures. %JC changed sentence from passive to active.  Or did somebody else reduce the data?

\begin{deluxetable*}{lcc|cccc}
\tablecaption{VLA and ALMA Continuum Properties \label{tbl-2}}
\tabletypesize{\scriptsize}
\tablewidth{0pt}
\tablehead{
\colhead{Source}  & \colhead{R.A.} & \colhead{Decl.} & 
\colhead{$S_{P}$} & \colhead{$S_{I}$} & \colhead{$\theta_{\rm M}\times\theta_{\rm m}$} & \colhead{$T_{\rm b}$}\\
\colhead{} & \colhead{(J2000)} & \colhead{(J2000)} &
\colhead{(mJy\,bm$^{-1}$)} & \colhead{(mJy)} & & \colhead{(K)}
}
\multicolumn{7}{c}{33\,GHz Components}\\[2pt]
\hline
\\[-6pt]
           NGC\,3627   &$   11   ~20   ~15.01   $&$+   12   ~59   ~29.5$  &     $1.70\pm0.06$  &     $2.09\pm0.08$  &                            $1\farcs21\pm0\farcs15 \times 0\farcs97\pm0\farcs10$  &     $1.98\pm0.34$\\
  NGC\,3627~Enuc.\,1   &$   11   ~20   ~16.30   $&$+   12   ~57   ~49.1$  &     $0.77\pm0.03$  &     $1.66\pm0.09$  &                            $2\farcs52\pm0\farcs23 \times 2\farcs25\pm0\farcs27$  &     $0.33\pm0.05$\\
  NGC\,3627~Enuc.\,2   &$   11   ~20   ~16.46   $&$+   12   ~58   ~43.5$  &     $0.98\pm0.04$  &     $5.35\pm0.27$  &                            $6\farcs95\pm0\farcs30 \times 3\farcs12\pm0\farcs17$  &     $0.28\pm0.02$\\
\hline
\\[-4pt]
\multicolumn{7}{c}{94.5\,GHz Components}\\[2pt]
\hline
\\[-6pt]
           NGC\,3627   &$   11   ~20   ~15.02   $&$+   12   ~59   ~29.6$  &     $1.22\pm0.07$  &     $1.58\pm0.11$  &                            $1\farcs10\pm0\farcs16 \times 0\farcs77\pm0\farcs29$  &   $0.252\pm0.102$\\
  NGC\,3627~Enuc.\,1   &$   11   ~20   ~16.29   $&$+   12   ~57   ~49.2$  &     $0.63\pm0.05$  &     $1.13\pm0.11$  &                            $1\farcs95\pm0\farcs24 \times 1\farcs24\pm0\farcs22$  &   $0.063\pm0.015$\\
  NGC\,3627~Enuc.\,2   &$   11   ~20   ~16.44   $&$+   12   ~58   ~43.5$  &     $0.68\pm0.05$  &     $3.23\pm0.25$  &                            $5\farcs66\pm0\farcs36 \times 1\farcs85\pm0\farcs15$  &   $0.042\pm0.005$  
\enddata
\end{deluxetable*}

\subsection{ALMA Band-3 Data}

In December 2013 as part of ALMA's Cycle 1 observing campaign, we obtained data for a subset of 9 SFRS 
sources %JC deleted '. These were'
 chosen to be sufficiently bright at 24\,$\mu$m and close enough on the sky that multiple sources could be observed in a single scheduling block. 
The array %JC inserted 'array'
configuration was chosen to match the $\lesssim$2\arcsec synthesized beam of the VLA 33\,GHz data. 
While the goal of the ALMA observing program is to deliver continuum images %JC replaced 'imaging'
 at $\approx$95\,GHz, we set the local oscillator frequency to 94.5\,GHz, centering
 the four 1.875\,GHz wide spectral windows at 87.5, 89.5, 99.5, and 100.5\,GHz, to 
cover the rest frequencies of %JC replaced 'allow for the potential detection of molecular species such as:'
 HCN ($J=1\rightarrow0$)/88.6318\,GHz, HCO$^{+}$ ($J=1\rightarrow0$)/89.1885\,GHz, HNC ($J=1\rightarrow0$)/90.6636\,GHz, and H40$\alpha$/99.0229\,GHz.  %JC rest frequencies with increased precision.  Note the recombination line at 99 GHz is H40alpha, not H30alpha.  H30alpha is at 231 GHz.  
Like the VLA data, the ALMA data were reduced and calibrated using CASA following standard procedures as part of the ALMA quality assurance process.
Here we present the nucleus and two extranuclear star-forming region in NGC\,3627. 
The remaining ALMA sources, along with corresponding VLA data, will be presented together in a forthcoming paper.  

\subsection{Archival CO ($J=1\rightarrow0$) Data}
To investigate how the amount of dense star-forming gas compares to the total molecular gas reservoir, we make use of $J=1\rightarrow0$ CO data taken as part of the BIMA SONG survey \citep{bimasong}.    
The rms noise of the CO channel map is 41\,mJy\,bm$^{-1}$ in a 10\,km\,s$^{-1}$ channel.  
The resolution of the CO map is significantly coarser than our ALMA and VLA data, having a synthesized beam of $7\farcs3 \times 5\farcs8$.

%, version 4.2;

%with used a spectral setup having 1.875\,GHz wide spectral windows centered at 87.5, 89.5, 99.5, and 100.5\,GHz to allow for the potential detection of HCN, HNC, and HCO$^{+}$.  
%We  set the LO frequency to be at 94.5\,GHz (i.e., the frequency between HCO$^{+}$(1-0)/89.2\,GHz and H30$\alpha$/99.0\,GHz, placing the centers of the four spectral windows at 87.5 and 89.5\,GHz in the LSB and 99.5 and 101.5\,GHz in the USB.  
%This observational setup  will cover the lines of HCN (1-0)/89.1\,GHz, HCO$^{+}$(1-0)/89.2\,GHz, HNC (1-0)/90.7\,GHz, CS(2-1)/98.0\,GHz, and H30$\alpha$/99.0\,GHz, among others.   
%This observational setup  will cover the lines of HCN (1-0)/89.1\,GHz, HCO$^{+}$(1-0)/89.2\,GHz, HNC (1-0)/90.7\,GHz, and H30$\alpha$/99.0\,GHz, among others.   

\subsection{Continuum Imaging and Photometry \label{sec:contphot}} %JC here 'imaging' is correct because this subsection describes how the images were created

Calibrated VLA and ALMA measurement sets for each source were imaged
using the task {\sc clean} in CASA.  The Ka-band images %JC replaced 'imaging' 
contain %JC now plural
data from both sets of observations, but are %JC now plural
 heavily weighted by the 13A semester observations %JC plural
as those include significantly more data.  The
mode of {\sc clean} was set to multifrequency synthesis \citep[{\sc
    mfs};][]{mfs1,mfs2}.  We chose to use {\it Briggs} weighting with
{\sc robust=0.5}, and set the variable {\sc nterms=2}, which allows
the cleaning procedure to also model the spectral index variations on
the sky.
%which allows the cleaning procedure to use components with a variable spectral index.  
To help deconvolve extended low-intensity emission, %JC deleted comma, added hyphen
we took %JC past tense 
advantage of the multiscale clean option \citep{msclean,msmfs} in CASA, searching for structures with scales 
$\approx$1 and 3  times the FWHM of the synthesized beam. 
%We applied 
A primary beam correction was applied using the CASA task {\sc impbcor}
before analyzing the images.    %JC active construction because we, not the correction, analyzed the images
The primary-beam-corrected %JC hyphens inserted
continuum images at 33 and 94.5\,GHz for each of the three %JC replaced '3' 
targeted sources are shown in Figure \ref{fig:contimgs}.   
The synthesized beamwidths and rms noises %JC 'beamwidths', not 'beams'; also plural noises
of each image are given in Table \ref{tbl-1}.
We also note that the largest angular scales %JC plural 'scales' 
that the ALMA and VLA images %JC replaced 'imaging' 
should be sensitive to are %JC plural verb
 $\approx$25\arcsec~and 44\arcsec, respectively.  

To measure the integrated flux densities and source sizes from each field, we used %JC past tense
 the task {\sc imfit} in CASA 
to fit sources within a circular aperture having a radius of 15\arcsec. %JC rephrased.  Why such a large aperture? 
In Table \ref{tbl-2} we list the deconvolved source parameters from %JC replaced 'fitting results reported by'
{\sc imfit}, including the positions %JC replaced 'center'
 of the source components, peak brightnesses ($S_{P}$), integrated flux densities ($S_{I}$), 
deconvolved source sizes ($\theta_{\rm M} \times \theta_{\rm m}$), and corresponding brightness temperatures ($T_{\rm b}$).   
In addition to the errors reported by {\sc imfit}, the uncertainties on the peak brightnesses and integrated  flux densities include the contribution from image rms, as well as an absolute calibration uncertainty of 3\% at 33\,GHz \citep{pb13fcal} and 5\% at 94.5\,GHz (ALMA Cycle 1 Technical Handbook), %JC Handbook, not 'Handbood'
 which in fact dominates the uncertainties.  
Clearly for the case of Enuc.\,2, which is an elongated structure at the end of the bar, a 
single-component %JC inserted hyphen
 fit is not the most appropriate.  
%%We therefore focus on resolved maps for the discussion of our results.  
%However, for simplicity in our comparisons, we choose to adopt the fit for the analysis.  

\subsection{Line Imaging and Photometry}

We detected % past tense
 the  $J=1\rightarrow0$ lines of HCN and HCO$^{+}$ towards each targeted 
region in NGC\,3627.  
We did not detect H40$\alpha$ for any of the targeted regions, and our frequency coverage missed HNC ($J=1\rightarrow0$) in each case.  
%, and HThe other two spectral lines sometimes included in our frequency coverage, HNC ($J=1\rightarrow0$) and H40$\alpha$, were not detected.  
Similar to the continuum imaging, the line data were imaged using the task {\sc clean} in CASA.  
However, before the line images were created, %JC past tense
the data were first continuum subtracted using the CASA task {\sc uvcontsub}. 
The rms noise of the ALMA channel maps is $\approx$1.2\,mJy\,bm$^{-1}$ in a 10\,km\,s$^{-1}$ channel.  

\begin{figure}
\center
\epsscale{1.2}
\plotone{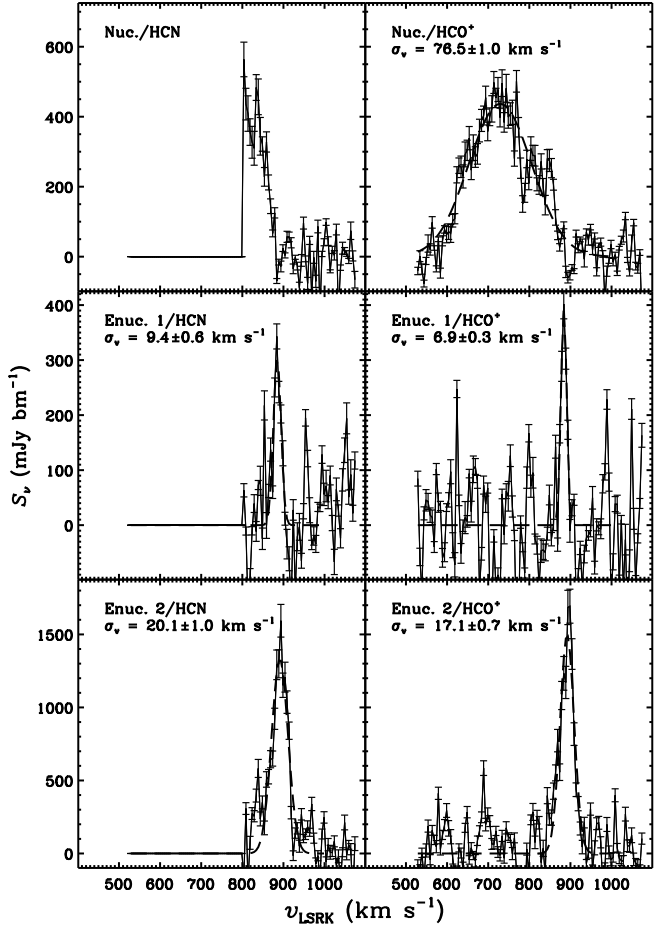}
\caption{
HCN and HCO$^{+}$ spectra (5\,km\,s$^{-1}$ channels) for each of the three targeted regions within NGC\,3627 extracted using a circular aperture centered on the 94.5\,GHz source positions returned by {\sc imfit}.  
The diameter of each aperture was 20\% larger than the corresponding source size major axis convolved with the 94.5\,GHz beam.  
The full velocity range of HCN emission towards the nucleus is cutoff by the bandpass at $v_\mathrm{LSRK} \approx 800\mathrm{\,km\,s^{-1}}$.  
For all other regions, the spectral lines are fit by Gaussians (dashed lines), and the corresponding fitted velocity dispersions and uncertainties are given in the upper left corner of each panel.  
}
\label{fig:vdisp}
\end{figure}

Moment 0 maps were constructed by integrating the spectra at each pixel of the spectral cubes.  
These are overlaid on the continuum maps in Figure \ref{fig:contimgs}.  
Unfortunately, the detection of HCN towards the nucleus is cut off %JC separated 'cutoff'
by the bandpass at a radial velocity of $v_\mathrm{LSRK} \approx 800\mathrm{\,km\,s^{-1}}$.  %JC is this a heliocentric velocity?  Specify.
The moment 0 map for HCO$^{+}$ towards the nucleus shows that we are missing higher velocity HCN emission to the north. %JC rephrased sentence
This can also be seen in Figure \ref{fig:vdisp}, which shows the spectra of both HCN and HCO$^{+}$ for all targeted regions.  
The spectra were extracted using a circular aperture centered on the 94.5\,GHz source positions returned by {\sc imfit} having a diameter that was 20\% larger than the corresponding source size major axis convolved with the 94.5\,GHz beam.  
We therefore exclude the nuclear HCN line emission in this study.  
For all other regions in Figure \ref{fig:vdisp}, the HCN and HCO$^{+}$ line profiles are fit by a Gaussian.  
The corresponding velocity dispersions are given the each panel, along with uncertainties that were estimated using the errors per channel along with a Monte Carlo approach.

Given that the CO ($J=1\rightarrow0$) image %JC the CO 'data' aren't at coarse resolution; the CO image is
has a much coarser angular resolution than %JC simpler phrase than 'is at such coarse resolution with respect to' 
our ALMA line images, %JC replaced 'imaging'
we must first
convolve our images %JC we convolved our images, not 'these data'  
to the resolution of the CO map before estimating
dense molecular gas fractions.  This must similarly be done for the
continuum (33 and 94.5\,GHz) data for estimating star formation
efficiencies (see \S\ref{sec:dgas-sfe}).  The resolution matching was
carried out using the task {\sc imsmooth} in CASA, before re-gridding
each image to a common pixel scale.  For such measurements requiring
the convolved data, photometry was carried out by simply taking %JC past tense
 the peak brightness within a circular aperture having a radius of 15\arcsec~(see Table \ref{tbl-3}), %JC radius? diameter? 
as this provides a way to mitigate any differences in our comparisons that may arise from being observed with different
interferometers having different sensitivities on different spatial scales. %JC rephrased final phrase

\begin{deluxetable*}{lcccc}
\tablecaption{Photometry Incorporating CO Data \label{tbl-3}}
\tabletypesize{\scriptsize}
\tablewidth{0pt}
\tablehead{
\colhead{Source}  & \colhead{$I_{\rm CO}$} & \colhead{$I_{\rm HCN}$} & \colhead{$I_{\rm HCO^{+}}$} & \colhead{$S_{\rm 33\,GHz}$}\\
\colhead{} & \colhead{(K\,Km\,s$^{-1}$)} & \colhead{(K\,Km\,s$^{-1}$)} & \colhead{(K\,Km\,s$^{-1}$)}& \colhead{(mJy\,bm$^{-1}$)}
}
\startdata
           NGC\,3627   &    375.47$\pm$ 19.3   &       4.54$\pm$0.48   &      13.80$\pm$0.81   &       2.10$\pm$0.06\\
  NGC\,3627~Enuc.\,1   &     48.19$\pm$  4.4   &       1.05$\pm$0.23   &       0.50$\pm$0.20   &       1.56$\pm$0.05\\
  NGC\,3627~Enuc.\,2   &    263.03$\pm$ 13.6   &       6.62$\pm$0.46   &       7.15$\pm$0.47   &       3.38$\pm$0.10  
\enddata
\tablecomments{Photometry was carried out after convolving both VLA and ALMA maps to the $7\farcs3\times5\farcs8$ sythesized beam of the BIMA SONG $J=1\rightarrow0$ CO data.}
\end{deluxetable*}

%JC Figure 3 is illegible because it is too small (especially the contour labels) and the deepest gray is nearly black, obliterating the central contours.

\begin{figure*}
\center
\epsscale{1.}
\plotone{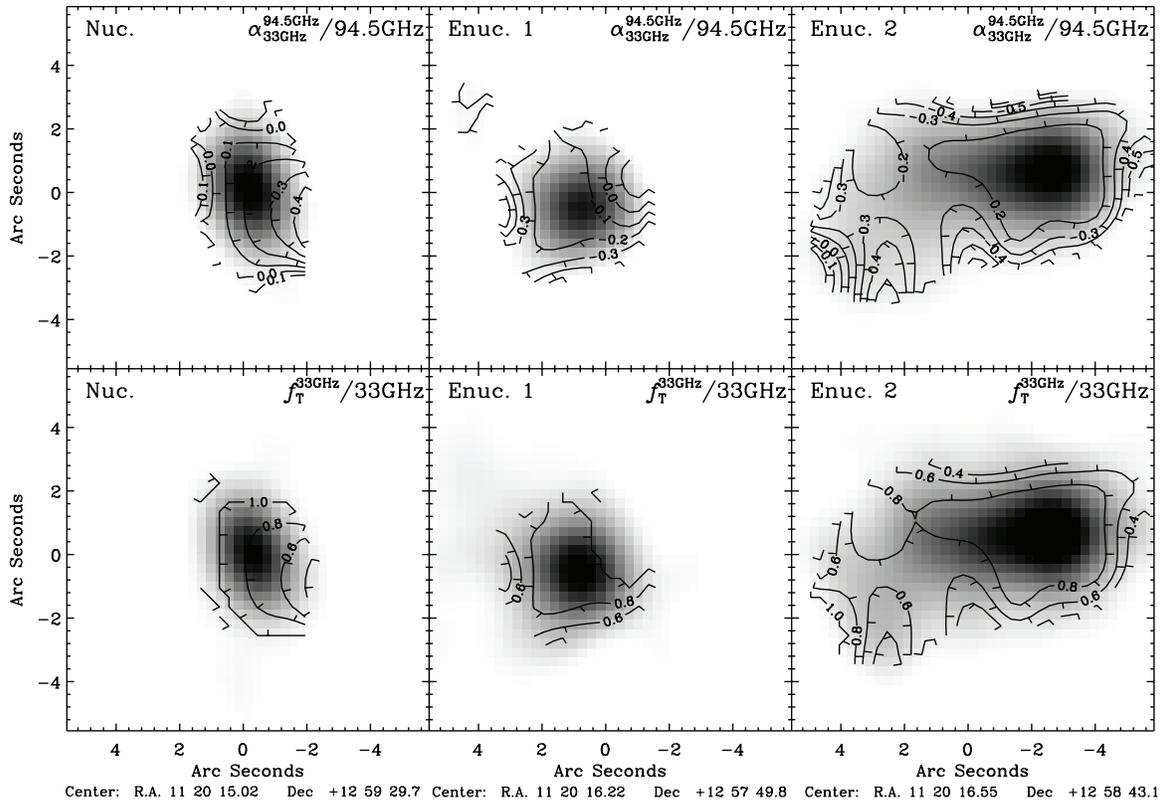}
\caption{
{\it Top row:} Spectral index contours between 33 and 94.5\,GHz overlaid on the 94.5\,GHz continuum image of the nucleus ({\it left}), Enuc.\,1 ({\it middle}), and Enuc.\,2 ({\it right}).   
%{\it Top Left:} Spectral index contours between 33 and 94.5\,GHz overlaid on the 94.5\,GHz continuum image of the nucleus of NGC\,3627.  
%{\it Top Middle:} The same as the top left panel, except for Enuc.\,1.  
%{\it Top Right:} The same as the top left panel, except for Enuc.\,2.  
In each case, the spectral indices flatten at the peaks of the continuum emitting regions.  
%There are also indications of spectral steepening at the outskirts of the star-forming regions, which may indicate the presence of thermal dust emission.  
%However, the uncertainties on the spectral indices become large in such regions.  
{\it Bottom row:} Estimated thermal fractions at 33\,GHz overlaid on the 33\,GHz continuum image of the nucleus ({\it left}), Enuc.\,1 ({\it middle}), and Enuc.\,2 ({\it right}) of NGC\,3627.   
%{\it Bottom Left:} Estimated thermal fractions at 33\,GHz overlaid on the 33\,GHz continuum image of the nucleus of NGC\,3627  
%{\it Bottom Middle:} The same as the bottom left panel, except for Enuc.\,1. 
%{\it Bottom Right:} The same as the bottom left panel, except for Enuc.\,2. 
For both extranuclear star-forming regions, the 33\,GHz thermal fractions approach unity on the peaks of the continuum emitting regions.  
}
\label{fig:spxtfrac}
\end{figure*}

\section{Results and Discussion}
In the following section we present our results along with a brief discussion about their 
implications for %JC replaced 'on' with 'for' 
the star-formation %JC inserted hyphen
activity in the three regions observed in NGC\,3627.

\subsection{Spectral Indices and Thermal Radio Fractions \label{sec:spxtfrac}}  %JC deleted estimates
We created %JC past tense
a 33-to-94.5\,GHz spectral-index %JC inserted hyphen
map %JC singular map for each source
 (i.e., $\alpha$, where $S_{\nu} \propto \nu^{\alpha}$) of each source after convolving the slightly higher resolution 94.5\,GHz data to match the beam of the 33\,GHz data and putting the data on the same grid.  
In the top panels of Figure \ref{fig:spxtfrac} the spectral indices are overlaid on 94.5\,GHz continuum images for which pixels not detected at the 3$\sigma$ level have been clipped.  
%for both extranuclear star-forming regions that have been clipped at the $3\sigma$ level.  %JC what has been clipped?
Focusing only on the regions (i.e., pixels) detected at $>3\sigma$ in each map, we measured %JC past tense
mean spectral indices %JC plural
of $\alpha_{\rm 33\,GHz}^{\rm 94.5\,GHz} = -0.10, -0.15,$ and $-0.27$ for the nucleus and extranuclear regions 1 and 2, respectively.  
%of $\alpha_{\rm 33\,GHz}^{\rm 94.5\,GHz} = -0.10 \pm 0.20, -0.15 \pm 0.18, -0.27 \pm 0.15$ for the nucleus and extranuclear regions 1 and 2, respectively.  %JC associated errors with measurements
%JC deleted 'The corresponding standard deviations are 0.20, 0.18, and 0.15, respectively.  
The corresponding standard deviations over these regions are 0.20, 0.18, and 0.15, respectively.  
As expected, these are %JC deleted 'found to be'
 relatively flat, being $< -0.5$ for each discrete source.  
We note that the spectral indices in some cases become positive in the outskirts of the H{\sc ii} regions, which may indicate a significant contribution of thermal dust to the 94.5\,GHz emission, although the uncertainties on the spectral indices in the  H{\sc ii} region outskirts %JC Does 'regions' here mean 'HII regions' or 'outskirts'?
become large.  

If we assume a fixed non-thermal spectral index for each source, we can use the measured spectral indices to estimate the
fractional %JC inserted 'fractional'
contributions %JC plural
 from thermal emission \citep[e.g., ][]{kwb84,ejm12b}.  
%We do not attempt this for the nucleus, as it is known to host an AGN for which there is likely to be more variation in the intrinsic non-thermal slope.   
While most sensitive to the assumption of the non-thermal spectral index, this simple thermal decomposition also assumes that the free-free emission does not become optically thick at $\nu \gtrsim 30$\,GHz \citep[e.g.,][]{tm10}, and that there is an insignificant contribution of anomalous microwave emission at $\sim$33\,GHz \citep[e.g.,][]{ejm10} and thermal dust emission at $\sim$94.5\,GHz.   
As discussed below, the few regions where there may be evidence for thermal dust emission are excluded from the thermal fraction calculations.   
%However, this calculation is most sensitive to the assumption of the non-thermal spectral index.}
We took %JC past tense
 the non-thermal spectral index to be $\alpha^{\rm NT} = -0.85$, which is %JC present tense
 the average non-thermal spectral index found among the 10 star-forming regions studied in NGC\,6946 by \citet{ejm11b} %JC deleted comma
 and very similar to the average value found by \citet[][i.e., $\alpha^{\rm NT} = -0.83$ with a scatter of $\sigma_{\alpha^{\rm NT}} = 0.13$]{nkw97} globally for a sample of 74 nearby galaxies.
The 33\,GHz thermal fraction contours are overlaid on the 33\,GHz continuum images in the bottom panels of Figure \ref{fig:spxtfrac}, indicating thermal fractions near unity on the peaks of the continuum emission.    

While NGC\,3627 is known to host an AGN, the AGN %JC 'it' is ambiguous here; is 'it' NGC 3627 or the AGN? 
does not appear to contribute significantly to the radio continuum emission.  
\citet{mef04} reported %JC past tense
 a 5\,GHz flux density of $< 0.3$\,mJy within a $3.6\,{\rm mas} \times 1.4\,{\rm mas}$ beam using the VLBA.    
Scaling this to 33\,GHz assuming a non-thermal spectral index of $\alpha^{\rm NT} = -0.85$ %JC The compact core might have a flat spectrum, so the VLBA core upper limit might be as high as 0.3 mJy (<10%) at 33 GHz
 results in an estimated AGN contribution to the 33\,GHz emission of $<$60\,$\mu$Jy , or $<$3\% of the integrated 33\,GHz emission for the nuclear component reported by {\sc imfit} (see \S \ref{sec:contphot}).  
We therefore assume that the nuclear emission is primarily %JC instead of 'entirely'
 powered by star formation in the rest of the analysis.    

Ignoring those regions for which the spectral index was measured to be flatter than $\alpha_{\rm T} = -0.1$ by more than 1$\sigma$, where %JC replaced 'since' by 'where'
 there may be a contribution from thermal dust, we measure corresponding mean 33\,GHz thermal fractions over the nucleus, Enuc.\,1 and 2 
of 85, 82 and 70\%, respectively.
% of $85 \pm 19$, $82 \pm 19$ and $70 \pm 20$\%, respectively.  
%JC deleted 'The corresponding standard deviations are 19, 19 and 20\%.' and associated errors with measured values
The corresponding standard deviations measured over these regions are 19, 19 and 20\%
The thermal fraction distribution and average values are in agreement with the results presented by \citet{ejm12b}, who reported an average 33\,GHz thermal fraction of $\approx$76\% with a dispersion of 24\% for their entire sample, and even higher values (i.e., $>$90\%, on average) for all sources resolved on scales of $\lesssim$500\,pc.  
We note that assuming a much flatter non-thermal spectral index has little effect on the estimated thermal fraction
 given the flat spectral indices measured between 33 and 94.5\,GHz.  
For example, by instead assuming $\alpha^{\rm NT} = -0.6$, the mean 33\,GHz thermal fractions over the nucleus, Enuc.\,1, and 2 are 80, 76, and 62\%, %JC inserted two commas
respectively, which are well within the quoted standard deviations.

\begin{figure}[!th]
\epsscale{1.1}
\plotone{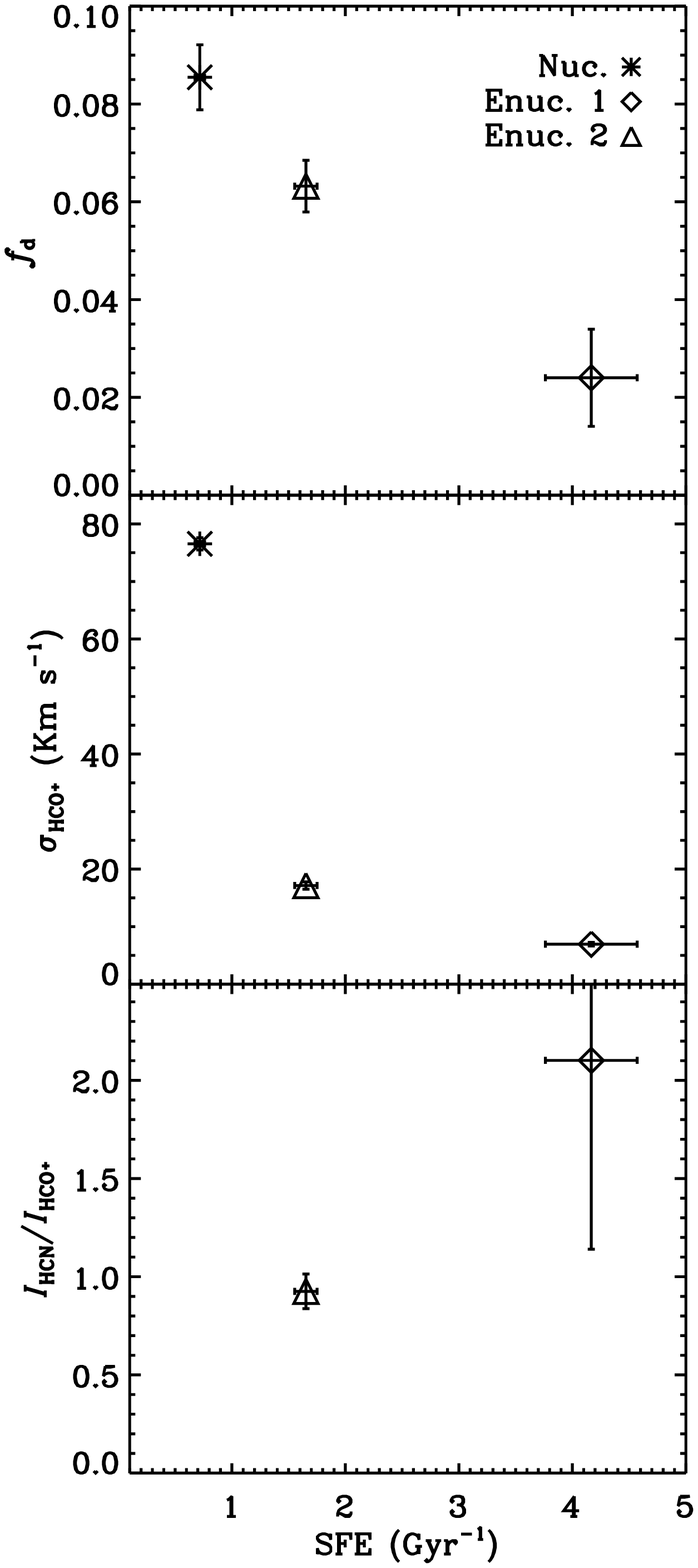}
\caption{{\it Top:} Dense gas fraction $f_{\rm d}$ plotted as a function of star formation efficiency for each region observed in NGC\,3627. 
The quantities on the two axes are both normalized by CO luminosity and so some spurious correlation is expected in the presence of noise. 
However, to the contrary we observe an anti-correlation between star formation efficiency and dense gas fraction, which is opposite the sense expected from only noisy CO measurements.  
{\it Middle:} HCO$^{+}$ velocity dispersions from the fitted line profiles shown in Figure \ref{fig:vdisp} plotted as a function of star formation efficiency for each region observed in NGC\,3627.  
{\it Bottom:} Ratio of HCN to HCO$^{+}$ brightnesses plotted against star formation efficiency, indicating a trend of increasing line brightness ratio with increasing measure of ionizing photons per unit molecular gas mass.  
Given that we are missing higher velocity HCN emission to the north of the nucleus, this data point is omitted in the bottom panel.  
%In both panels, measurements were made by taking the peak brightness within a 15\arcsec radius aperture after first convolving/re-gridding all maps to the resolution/pixel-scale of the $J=1\rightarrow 0$ CO data.  }
}
\label{fig:dgas}
\end{figure}

\subsection{Spatial Offsets Between Continuum and Line Emission \label{sec:offset}}  
One of the most obvious results from an inspection of the maps in Figure \ref{fig:contimgs} is the spatial offset between the peaks of the continuum and line emission.  
This is especially evident for Enuc.\,2 (i.e., the bar end) for which both 
dominant HCN and HCO$^{+}$ peaks are well to the east of the 33 and 94.5\,GHz continuum peaks.  
The continuum and molecular %JC inserted 'molecular' to distinguish from the ionized gas making the continuum
gas are spatially offset in the plane of the sky by $\approx$3\arcsec ($\approx$130\,pc at the distance of NGC\,3627).  
%Using the {\sc imfit} results, we the spatial offsets between the continuum and line emission for Enuc.\,2 is $XX\pm YY\arcsec$ ($ZZ\pm WW$\,pc).  

%HCN and HCO$^{+}$  are the two most abundant H$_{2}$ mass tracers after CO.  
The critical density of both the $J=1\rightarrow0$ lines of HCN and HCO$^{+}$ is $n_{\rm crit} \gtrsim 10^{5}$\,cm$^{-3}$, and thus probes the densest, UV-shielded gas that is likely in the process of, or soon will be, forming stars.  
Assuming that the 33 and 94.5\,GHz continuum is primarily powered by free-free emission, as suggested by the results presented in \S\ref{sec:spxtfrac}, and thus robustly traces %JC replaced 'tracing'
 ongoing star formation that is $\lesssim$10\,Myr old, %JC inserted 'old'
 this offset suggests that we have reached the spatial resolution at which we can reliably %JC reliably, not reliable
 separate the fuel stockpiles from the current generation of star formation.

\citet{pan13} similarly report $\sim$100\,pc offsets between peaks of dense gas (HCN $J=1\rightarrow0$) and star formation (3\,cm radio continuum) in the starburst ring of the barred galaxy NGC\,7522.  
These authors suggest that dense-gas %JC inserted hyphen
formation is promoted by gas cloud collisions at the intersections between the galaxy bar and ring orbits, and that offsets between the dense gas peaks and those of ongoing star formation occur farther %JC 'farther' not 'further' when discussing actual (not metaphorical) distances 
downstream from the orbit contact points.  
Similar conclusions have also been found for nuclear bars in nearby starbursts \citep[e.g.,][]{mth08}.  
Like NGC\,7522, NGC\,3627 is %JC deleted redundant 'also'
 a barred galaxy with an inner molecular ring \citep{mwr02}, suggesting that similar dynamics may also be driving the formation of dense gas and its observed spatial offsets from %JC replaced 'offset with'
 the peaks of ongoing star formation activity in Enuc.\,2.  

Assuming this scenario to be true, we can make a rough estimate of the propagation time between the newly formed stars, as traced by the free-free radio continuum emission, and the dense gas peaks.   
The rotation speed of the NGC\,3627's disk at the distance to the bar end (i.e., $\approx$100\arcsec~$\approx$4.5\,kpc) is $\approx$204\,km\,s$^{-1}$\citep{deBlok08}.    
The bar pattern speed for NGC\,3627 is $\approx$55\,km\,s$^{-1}$\,kpc$^{-1}$ \citep{kartik02}, corresponding to $\approx$247\,km\,s$^{-1}$ at this distance, yielding a net velocity of between the disk and the bar of $\approx$43\,km\,s$^{-1}$.  
Thus, the propagation time is $\approx$3\,Myr, 
less than the age of typical OB stars.  %JC 'much less than' or 'comparable with'?  O stars have short main-sequence lifetimes. see http://en.wikipedia.org/wiki/O-type_star

Another striking result from comparing the Enuc.\,1 maps in Figure \ref{fig:contimgs} is that the HCN and HCO$^{+}$ peaks are not co-spatial.  
The peak of the HCO$^{+}$ emission is located near the peak of the 33 and 94.5\,GHz continuum emission.  
However, the HCN emission peaks in regions that surround the HCO$^{+}$ and continuum peaks.  
This may be the result in variations in the (dense) gas density, given that the critical density of HCN is nearly and order of magnitude larger than that of HCO$^{+}$ \citep[e.g., ][]{mt12}.  
Stockpiles of ultra dense gas are likely still present in the immediate vicinity surrounding the newly formed H{\sc ii} region.  
Although, excitation effects may also be in play (e.g., see \S\ref{sec:dgas-sfe}).

 \subsection{Dense Gas Fractions and Star Formation Efficiencies \label{sec:dgas-sfe}}  

In the top panel of Figure \ref{fig:dgas} we plot the dense molecular gas fraction against the estimated star formation efficiency for each region.  
The dense molecular gas fraction is taken to be the mass ratio of dense molecular gas traced by the $J=1\rightarrow0$ transition of HCO$^{+}$ to the total molecular gas traced by the $J=1\rightarrow0$ transition of CO.  
For this calculation we assume HCO$^{+}$-to-H$_{2}$ and CO-to-H$_{2}$ conversion factors of $\alpha_{\rm HCO^{+}} \approx 10$\,M${_\odot}\,{\rm (K\,km\,s^{-1}\,pc^{2})}^{-1}$ \citep[e.g.,][we assume $\alpha_{\rm HCO^{+}} \approx \alpha_{\rm HCN}$]{gao04} and $\alpha_{\rm CO} \approx 4.3$\,M${_\odot}\,{\rm (K\,km\,s^{-1}\,pc^{2})}^{-1}$ \citep[e.g.,][]{svb05,bwl13}, respectively.  %JC overloading alpha to mean both spectral index and molecular gas conversion factors may confuse the reader.  The usual symbol for the CO to H2 conversion factor is X_CO.  Maybe X should be used instead of alpha here for all of the molecules.
We only make use of the HCO$^{+}$ for deriving the dense molecular gas masses as we have data for all regions.  
The star formation efficiency (SFE), whose inverse is often referred to as the gas depletion time, is taken to be the ratio of the star formation rate (SFR) to the total molecular gas mass traced by the $J=1\rightarrow0$ transition of CO.  
We converted %JC past tense
the 33\,GHz spectral luminosity into a star formation rate using the calibration for pure thermal emission given in \citep{ejm11b,ejm12b} and assumed %JC past tense
an average 33\,GHz thermal fraction of 79\% (see \S\ref{sec:spxtfrac}).  
%A clear trend of increasing star formation efficiency with decreasing dense gas fraction is observed.  
A clear trend of decreasing dense gas fraction with increasing star formation efficiency is observed.  
More specifically, the nuclear disk of NGC\,3627 appears to have the largest fraction of its molecular gas in a dense phase, 
%the largest reservoir of dense molecular gas, 
followed by the star-forming region at the end of the bar (Enuc.\,2), and the isolated H{\sc ii} region (Enuc.\,1). 
However, Enuc.\,1 (the isolated H{\sc ii} region), appears to be converting its molecular gas into stars more efficiently %JC 'efficient rate' doesn't make much sense.  Efficiency is a dimensionaless quantity, while star-formation rate has dimenssions of solar masses per year. Replace 'at a more efficient rate' with 'at a higher rate'?
(i.e., $\approx$ a factor of 3 more rapidly) than either %JC use 'either' with 'or', or 'both' with 'and'
the star formation region at the end of the bar or the nuclear disk.  
By plotting the HCO$^{+}$ velocity dispersions from the fitted line profiles in Figure\,\ref{fig:vdisp} against star formation efficiency in the middle panel of Figure \ref{fig:dgas}, we find that 
%Looking at the HCO$^{+}$ line profiles in Figure\,\ref{fig:vdisp}, we find that 
the star formation efficiency for each region increases with decreasing velocity dispersion.  
Thus, perhaps unsurprisingly, the dynamical state of the dense gas appears to have a larger impact on the star formation process than the actual fraction of dense gas that is available for star formation.  

Comparing the dense gas fraction and the efficiency of star formation on much larger scales, \citet{usero15} recently found systematic trends in both the dense gas fraction and the efficiency with which dense gas forms stars. 
That study included NGC\,3627, though with a beam area two orders of magnitude larger than our study. 
Similar to our results, but using HCN, they found that in $\approx 30$ disk galaxies, the apparent dense gas fraction increases by moving from the outer, low surface density parts of galaxies to the inner regions. 
At the same time, they found the apparent efficiency with which dense gas forms stars (for them, SFR/HCN) appears to decrease as one moves from the lower surface density disk to the inner parts of galaxies. 
They showed that these results could be explained by models of turbulent clouds \citep[e.g.,][]{km05,fk12} in which the average density and the turbulent Mach number in a cloud affect both the dense gas fraction and the ability of gas at different densities to form stars. 
Similar results have been found considering the apparently low rate at which dense gas forms stars in the central part of the Milky Way \citep[e.g.,][]{longmore13, jkauffmann13, rathborne14}
With the improved resolution offered by ALMA, we are able to measure a velocity dispersion that may be more directly related to the turbulent velocity dispersion (at least outside the central region). 
The observation of variable dense gas efficiency (SFR/HCO$^{+}$) and suppressed star formation in the central, high dispersion part of the galaxy appears to agree at least qualitatively with these results.  

Since we have relied on only %JC moved 'only' for correct logic
the $J=1\rightarrow0$ transition of HCO$^{+}$ to derive the dense gas mass, and have assumed similar conversion factors for HCN and HCO$^{+}$ to dense H$_{2}$, it is illustrative to see how the HCN-to-HCO$^{+}$ line brightness ratio varies among %JC use 'between' for two regions, 'among' for three or more retions
regions.  
This is shown in the bottom panel of Figure \ref{fig:dgas}, and plotted against star formation efficiency.  
The HCN-to-HCO$^{+}$ line brightness changes by a factor of $\approx$2 between both extranuclear star-forming regions over a similar change in star formation efficiency.  
This trend may %JC what probability does 'may likely' indicate?  the modest 'may'? the high 'is likely'?
be due to gas excitation effects given that the abscissa is a measure of the number of ionizing photons 
(free electrons) %JC deleted hyphen 
per unit molecular gas mass.  
For instance, as discussed in \citet{ppp07}, the HCO$^+$ abundance is known to be sensitive to the ionization 
degree %JC do you mean ionization degree (number of electrons removed from each molecule) or ionization fraction?
of molecular gas, which can significantly reduce the HCO$^+$ abundance in star-forming and highly turbulent molecular gas, while HCN remains abundant.  
Given the observed trend, this appears to be a plausible explanation.  
However, there are many alternative explanations for changes in the ratio of HCN-to-HCO$^{+}$ line brightness ratio including variations in gas density \citep[see, e.g., ][]{mt12}.  
%It is possible that this could be due to a gas excitation effect, given that the abscissa is a measure of the number of ionizing photons (free-electrons) per unit molecular gas mass.  
%The increase in free electrons could result in few  , 
%However, the error bar on the line-brightness ratio is simply too large for Enuc.\,1 to warrant further speculation.   
Given the large error bar on the line-brightness ratio for Enuc.\,1, further speculation here is unwarranted.  
Regardless, it is worth pointing out that even if one ignores the large error bar, the trend observed in the top panel of Figure \ref{fig:dgas} would persist even if we instead used the HCN-derived dense gas mass.

\section{Conclusions}
In this paper we have combined ALMA/Band-3 line and continuum images %JC replaced 'imaging'
 with VLA/Ka-band images %JC not 'data'
 to characterize the star formation activity on $\approx$100\,pc scales around three distinct regions within the nearby galaxy NGC\,3627; i.e., a nuclear disk hosting an AGN, an isolated H{\sc ii} region (Enuc.\,1), and a star-forming complex at the end of a bar (Enuc.\,2).  
Our conclusions can be summarized as follows:

\begin{itemize}
\item 
%Using the VLA 33\,GHz and ALMA 94.5\,GHz continuum data, we have constructed spectral index and thermal fraction maps around the nucleus and two extranuclear star-forming regions in NGC\,3627. 
The thermal fraction at 33\,GHz is nearly %JC replaced 'largely' with 'nearly'
 unity at the peaks of the H{\sc ii} regions as mapped on %JC deleted extra 'on'
 $\approx$100\,pc scales, %JC inseted 'scales'
with an average value of $\approx$76\% for both extranuclear star-forming regions.  
The mean thermal fraction at 33\,GHz among the three regions studied is 79\% with a dispersion of 19\%.  
We additionally found %JC past tense
the radio spectral index to become both increasingly negative and positive 
% negative and positive spectral steepening %JC ambiguous; use a different term
away from the peaks of the  H{\sc ii} regions, indicating an increase of extended  %JC 'extended' not 'diffuse'
non-thermal emission from diffusing cosmic-ray electrons and the possible presence of cold dust, respectively.  

\item The peaks of the $J=1\rightarrow0$ HCN and HCO$^{+}$ line emission are spatially offset from the peaks of the 33 and 94.5\,GHz continuum emission.  
For Enuc.\,2, the continuum and gas are spatially offset in the plane of the sky by $\approx$3\arcsec ($\approx$130\,pc at the distance of NGC\,3627).  
Assuming that the 33 and 94.5\,GHz continuum is primarily powered by free-free emission, and that the $J=1\rightarrow0$ HCN and HCO$^{+}$ line emission is tracing dense, UV-shielded gas, this indicates that our data reach an angular resolution at which one can spatially distinguish sites of current star formation from the fuel stockpiles for the next-generation of star formation.  

\item Combining our ALMA and VLA observations with archival BIMA CO ($J=1\rightarrow0$) data, we calculate dense gas fractions and star formation efficiencies for each region, finding that the dense gas fraction decreases with increasing star formation efficiency.  
This suggests %JC 'may suggest' is pretty weak.  If we can's just say 'suggest', maybe we shouldn't say anything.
that an increase in the dense gas content of star-forming regions does not reflect an increased efficiency for which parts of galaxies can turn molecular gas into stars among the three, diverse regions studied here.  
Specifically, Enuc.\,1 (an isolated H{\sc ii} region) %JC deleted comma after )
 appears to be converting molecular gas into stars more efficiently than both the star-forming region at the end of the bar or the nuclear disk.  
We additionally find that the velocity dispersion of the dense gas in each region decreases with increasing star formation efficiency, indicating that the dynamical state of the dense gas, rather than its abundance, plays a larger role affecting ongoing star formation activity.  

\item The $J=1\rightarrow0$ HCN/HCO$^{+}$ brightness ratio and star formation efficiency both vary %JC replace 'vary' by 'differ'? 
by a factor of $\approx$2 between the extranuclear H{\sc ii} regions.  
We speculate that this may be due to an increase in the ionization degree %JC ionization 'degree' or 'fraction'?
 of molecular gas for the star-forming region having a higher star-formation %JC inserted hyphen
efficiency, which in turn reduces the HCO$^+$ abundance while HCN remains abundant.
However, alternative explanations (e.g., variations in the gas density) may also be plausible.

\end{itemize}

\acknowledgements
We would like to thank the anonymous referee for useful comments that helped to improve the content and presentation of this paper.  
%This work was supported in part by National Science Foundation Grant No. PHYS-1066293 and the hospitality of the Aspen Center for Physics. 
%E.J.M. acknowledges the hospitality of the Aspen Center for Physics, which is supported by the National Science Foundation Grant No. PHY-1066293.  
The National Radio Astronomy Observatory is a facility of the National Science Foundation operated under cooperative agreement by Associated Universities, Inc.
This paper makes use of the following ALMA data: ADS/JAO.ALMA\#2012.1.00456.S. 
ALMA is a partnership of ESO (representing its member states), NSF (USA) and NINS (Japan), together with NRC (Canada) and NSC and ASIAA (Taiwan), in cooperation with the Republic of Chile. 
The Joint ALMA Observatory is operated by ESO, AUI/NRAO and NAOJ. 
%This research has made use of the NASA/IPAC Infrared Science Archive, which is operated by the Jet Propulsion Laboratory, California Institute of Technology, under contract with the National Aeronautics and Space Administration.
This research has made use of the NASA/IPAC Extragalactic Database (NED), as well as the NASA/IPAC Infrared Science Archive, both of which are operated by the Jet Propulsion Laboratory, California Institute of Technology, under contract with the National Aeronautics and Space Administration.

%\bibliography{/Users/emurphy/libs/bibtexref/master_ref}
\bibliography{aph.bbl}

\end{document}